# Brownian reservoir computing realized using geometrically confined skyrmions

**Author**

Klaus Raab[1], Maarten A. Brems[1], Grischa Beneke[1], Takaaki Dohi[1], Jan Rothörl[1], Johan H. Mentink[2,*], Mathias Kläui[1,3,*]

**Affiliation**

[1] Institut für Physik, Johannes Gutenberg-Universität Mainz, Staudingerweg 7, 55128 Mainz, Germany

[2] Radboud University, Institute for Molecules and Materials, Heyendaalseweg 135, 6525 AJ Nijmegen, The Netherlands

[3] Graduate School of Excellence Materials Science in Mainz, Staudingerweg 9, 55128 Mainz, Germany

**\*E-mail addresses (Corresponding authors)**

Klaeui@uni-mainz.de

J.Mentink@science.ru.nl

**Abstract**

Reservoir computing (RC) has been considered as one of the key computational principles beyond von-Neumann computing. Magnetic skyrmions, topological particle-like spin textures in magnetic films are particularly promising for implementing RC, since they respond strongly nonlinear to external stimuli and feature inherent multiscale dynamics. However, despite several theoretical proposals that exist for skyrmion reservoir computing, experimental realizations have been elusive until now. Here, we propose and experimentally demonstrate a conceptually new approach to skyrmion RC that leverages the thermally activated diffusive motion of skyrmions. By confining the electrically gated and thermal skyrmion motion, we find that already a single skyrmion in a confined geometry suffices to realize non-linearly separable functions, which we demonstrate for the XOR gate along with all other Boolean logic gate operations. Besides this universality, the reservoir computing concept ensures low training costs and ultra-low power operation with current densities orders of magnitude smaller than those used in existing spintronic reservoir computing demonstrations. Our proposed concept can be readily extended by linking multiple confined geometries and/or by including more skyrmions in the reservoir, suggesting high potential for scalable and low-energy reservoir computing.

# Main text

**Introduction**

Skyrmions are magnetic whirls with topologically enhanced stability, which can behave like two-dimensional quasi-particles[1,2]. Skyrmions have been found in thin metal films[3–7] and bulk materials[8] and the efficient displacement due to spin-transfer-torques[9] and spin-orbit-torques (SOT)[10–14] by low current densities bears tremendous potential for non-conventional computing[15–20] and novel types of memory devices[21,22].

While deterministic skyrmion dynamics due to current-induced torques have previously been exploited[12], recently, it has been shown that skyrmions exhibit stochastic dynamics induced by thermal diffusion[17]. This dynamics is intrinsically nonlinear as well and moreover can be easily influenced by geometrical confinement[23,24], where the equilibrium configurations of skyrmion arrangements can be controlled by geometry. In particular, by exploiting commensurability effects, the ordering of the equilibrium arrangements can be tailored[24]. Furthermore, the current-induced torques have so far not been systematically combined with thermally excited diffusion but potentially this combination can reduce the current densities for current-induced motion dramatically. Exploiting skyrmion motion for nonconventional computing[25] has been considered highly promising due to the combination of both intrinsic nonlinear dynamics and stochastic dynamics, which are both features of the human brain as well[26].

One of the key paradigms in brain-inspired computing is reservoir computing (RC)[27], which leverages the nonlinear dynamics of a medium to map a complex problem to a much simpler linear problem. Since the reservoir itself does not need to be trained, training costs reduce to that of solving a linear problem, yielding fast and low-energy learning[28]. Although several theoretical proposals exist for implementing RC with magnetic skyrmions, they heavily rely on the existence of local pinning sites[15,29], and the nonlinear dynamics features small displacements making readout challenging experimentally. Moreover, for extended magnetic skyrmions textures, reproducible operation requires an external reset mechanism. On top of that, the skyrmion dynamics relying on deterministic motion induced by high current densities[17] limits the potential to improve the energy efficiency as compared to existing spintronic RC[30].

We overcome these challenges by designing and experimentally realizing a conceptually new approach: a Brownian skyrmionic device as a RC component, which exploits the intrinsic properties of thermally active skyrmions in geometrical confinement combined with ultra-low power current-induced dynamics. An effective potential well created by the confinement allows for a natural reset mechanism, which does not rely on pinning effects, but is instead enabled by the thermal fluctuation of the skyrmions in combination with the geometrical confinement itself. To demonstrate the functionality of this RC device, we exemplify reliable and reproducible Boolean logic operations including the nonlinearly separable XOR operation.

**Functionality of the device**

Fig. 1 schematically shows the system under consideration, in which a confinement geometry of an equilateral triangle harbors a single skyrmion (sample and fabrication details are given in the methods section). This simple and highly symmetric geometry allows for a variety of different states (skyrmion positions) depending on the voltages applied to the contacts at the three corners of the triangular confinement. The device is channeled with narrow wires at the tips for better connectivity to the gold pads. When voltage potentials are applied to the contacts, the skyrmion position depends on the interplay between current-induced motion due to SOTs[10–14], skyrmion-edge repulsion[23,24] and thermal diffusive dynamics[17]. The state of the system is imaged by magneto-optical Kerr-effect microscopy (imaging setup is described in the methods section). To evaluate the performance of the system as part of a RC device, we track the skyrmion positions and mimic read-out via magnetic tunnel junctions (MTJs) as described below. In principle, many input combinations are possible, however, to resemble Boolean functions below, we restrict ourselves to grounding one contact and employing ground or positive voltage values as inputs 0 or 1 respectively at the other contacts (Fig. 1).

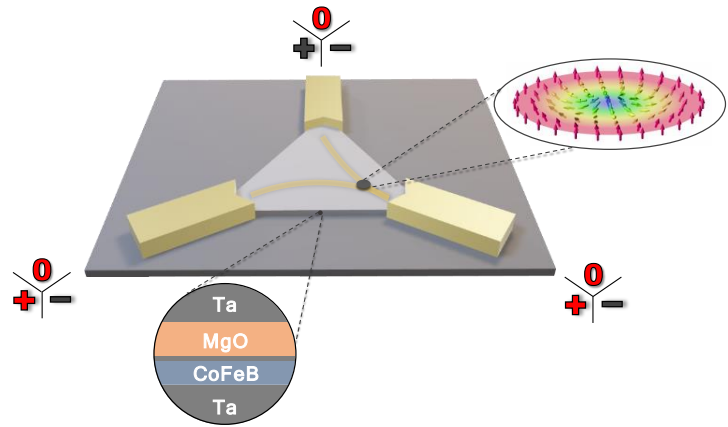

Fig. 1| **Three-dimensional schematic of the device.** The stack structure is shown in the lower circle. Attached are the chromium/gold contacts, on which positive, negative, or null potential can be applied. The input values used for the present Boolean logic demonstration are highlighted in red. The dark grey spot in the triangle represents a skyrmion as imaged in our MOKE recordings. Here, it is pushed into the lower right corner (yellow lines represent schematically the current flow). The inset on the right depicts the schematic spin structure of a Néel-skyrmion.

The skyrmions are nucleated as described in a previous work[17] by applying an in-plane magnetic field pulse on top of a static out-of-plane field at the spin reorientation transition. One of the main advantages of this device is the automatic initialization: by, ideally, nucleating a single skyrmion in the device, the system immediately relaxes to the required ground state as the skyrmion resides primarily in the central region of the confinement due to skyrmion-edge repulsion[23,24] as shown in Fig. 2a). Thereby and in particular without the necessity of further input or adjustment of the local pinning effect[15,29], the ground state, in which the system resets itself when no potential is applied, is achieved. Thus, the interplay of thermal dynamics and edge repulsion acts as an auto-initialization and reset mechanism simultaneously.

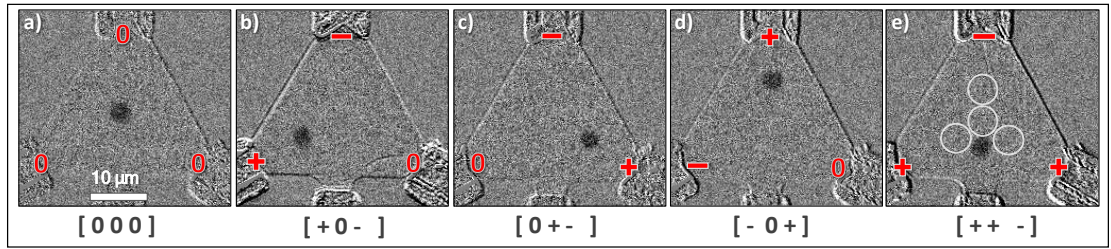

Fig.2 | **Skyrmion displacement.** Subtracted Kerr microscopy images of an equilateral triangular device with edge length 36 µm (tip to tip), each with a single skyrmion (dark grey) in a) the ground state without and b)-e) with applied electrical potentials. In brackets below the input patterns of the respective state. e) shows exemplary the four regions used to mimic MTJs in our analysis. The chromium/gold contacts overlap the thin film in the corners and on the bottom edge (the latter is unused for this publication).

**Current induced skyrmion dynamics**

As seen in Fig. 2a) the skyrmion without any applied voltage rests in the middle region of the triangle. For example, by applying a positive potential at one corner, a negative at an opposite corner and ground (null potential) to the remaining one, we observe the current-induced motion of the skyrmion towards the corner with the applied positive potential. We observe current-induced motion starting at $5\times10^7$ A m$^{-2}$ in a direction opposite to the direction of the current flow, which is consistent with previous work but at a current density that is about four orders of magnitude smaller[17]. This drastically reduced threshold current density allows for ultra-low power operation. As the current density is increased, the effects of the torques become more and more dominating until increasing the current density further, eventually results in pushing the skyrmion into one of the corners, where the current density is highest (see supplementary material) and the skyrmion annihilates. Here we choose a current density in the range of $10^8$ A m$^{-2}$, which is sufficient for reliable and quick operation while not resulting in any skyrmion annihilation events.

Since the current density increases monotonically towards the tip of the triangle element, we calculate the current density of the device at the half-width of the triangle at the corner from which the current originates. While we have a geometrically symmetric system, the resistance between the pairs of contacts/corners slightly differs due to fabrication irregularities. Hence, the same voltage potential leads to different current density values, which, additional to possible pinning sites, influences the skyrmion occurrence. A deviation from the ideal equilateral triangle does not impede operation, since the thermally activated diffusive dynamics allows the skyrmions to explore the full state space. Such irregularities are accounted for by default in the training of the linear read-out (see detailed discussion in the supplementary material). Hence, we demonstrate that our RC method automatically compensates for sample/setup imperfections.

**Evaluation and linear regression**

The general approach is to manipulate the skyrmion dynamics via the applied electric potentials and measure the probability for the skyrmion to be present within different regions of the sample. For now, the read-out is performed optically by imaging and tracking the skyrmions. For a scaled-down device, this could, for instance, be technically done using average tunnel magnetoresistance (TMR) as the TMR depends on the presence of a skyrmion in the relevant region[31]. To mimic read-out via MTJs[31], we employ four circular read-out regions with a radius of 2,2 µm within the confinement arranged center-symmetrically as indicated in Fig. 2e). The images in Fig. 2b)-e) show only exemplarily the displacement of a single skyrmion from the ground state in Fig. 2a). Note, that we have studied a number of devices that all show qualitatively the same behavior. The results shown below stem from a device with 40 µm edge length.

The local skyrmion occurrence probabilities within the four circular regions are processed externally via linear read-out, i.e., the output is given by the weighted sum of the probabilities plus an offset. Thereby, the same device can perform a multitude of different operations depending on the weights.

The skyrmion occurrence is measured using Kerr-microscopy for 13,000 frames at 16 frames per second. To obtain a series of local occurrence probabilities for the four regions, we average the skyrmion occurrence over 62.5 s time intervals. For each input combination we use the first 4 of the resulting 13 sets of local occurrences probabilities to optimize (train) the weights for linear read-out using the Scikit-learn software package[32]. The Output $Q$ is then given by

$$Q = W_{\text{left}}P_{\text{left}} + W_{\text{right}}P_{\text{right}} + W_{\text{top}}P_{\text{top}} + W_{\text{middle}}P_{\text{middle}} + W_{\text{intercept}}, \quad (1)$$

where $W_{\text{region}}$ is the weight of the probability $P_{\text{region}}$ for the skyrmion to be in the specific circular region and $W_{\text{intercept}}$ is an offset. The probability $P_{\text{region}}$ itself depends on the input patterns of the device ([0 0], [0 1], [1 0], [1 1]) (see Fig. 3).

Fig. 3 shows the output of the linear read-out, trained for the Boolean operations AND, NAND, OR, NOR, XOR, and XNOR. The training-set is indicated as the light blue part of the curves whereas the test-set is black. For all Boolean operations we observe values close to the expected output and a good separation even on the test-set. The corresponding optimized weights can be found in the supplementary material (supplementary table 1). The dashed lines show possible thresholds for perceptron read-out, i.e., values above and below the threshold are assigned to 1 and 0, respectively. In particular, we have demonstrated NAND and NOR functionality, each representing a functionally complete set of logical connectives, and the non-separable XOR functionality. The latter demonstrates that this system with just one confined skyrmion is already sufficiently complex to perform non-linearly separable tasks, which is impossible for instance using a conventional single-layer perceptron readout alone.

The signal to noise ratio ($SNR$) of the read-out is defined as

$$SNR := \frac{\langle T \rangle - \langle F \rangle}{\sigma_T + \sigma_F}, \qquad (2)$$

where $T$ and $F$ are the sub-sets of linear read-out outputs $Q$, for which the corresponding Boolean operation applied to the device input patterns give True or False, respectively. Angled brackets indicate the mean value and $\sigma_T$ ($\sigma_F$) is the standard deviation of the $T$ ($F$) subset. Averaged over the six different Boolean operations, the data in Fig. 3 exhibits $SNR > 5$.

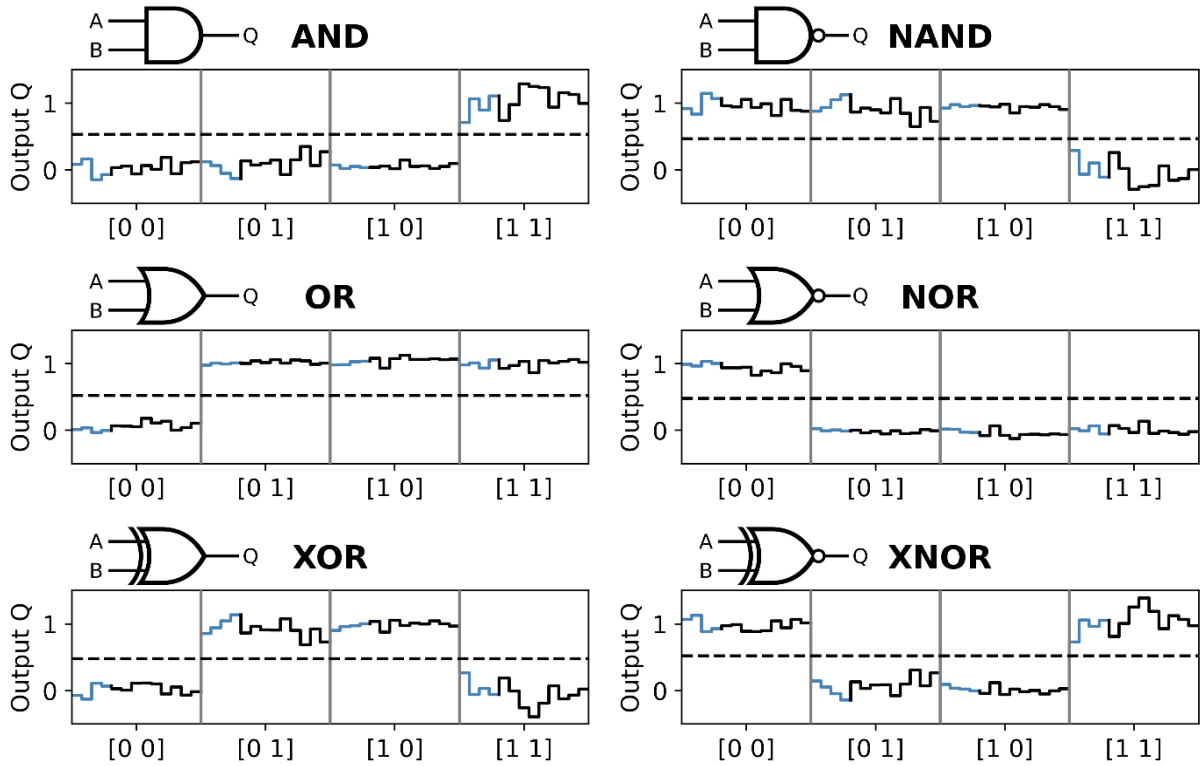

Fig. 3| **Logic operations.** Outputs of the linear read-out optimized for different Boolean operations. For each input combination ([0 0], [0 1], [1 0], [1 1]) the output Q of the linear read-out is shown for 13 sets of local skyrmion occurrence probabilities. The light blue and black parts of the curves indicate the sets used for training and testing, respectively. The dashed horizontal line indicates a possible threshold for perceptron read-out.

The $SNR$ decreases as the time interval used to obtain the local skyrmion occurrence probabilities is decreased. For example, if the time interval is divided by two while the ratio between the number of sets used for training and testing is kept constant, the $SNR$ decreases to $SNR \approx 4$. Note here that the actual timescale of skyrmion diffusion is much faster than the chosen time intervals[17] and even faster for smaller devices as the characteristic time scale of directed diffusive motion scales with the square of the corresponding length scale[33]. Thus, we mainly attribute this effect on the employed time scales to the low sampling rate of the Kerr microscope.

**Conclusion**

While the presented Brownian RC already suffices to realize nonlinearly separable operations for a single confined skyrmion, we emphasize that the concept can easily be generalized by exploiting more input combinations and by increasing the number of skyrmions in the geometry. With higher numbers of skyrmions, the system exhibits more complex dynamics like (un-)commensurable ground states[24] and thus enhances the accessible states for the RC. Moreover, the complexity can be further enhanced by linking multiple confined geometries, which could provide an ultralow-energy alternative to neuromorphic computing based on arrays of nanoscale spintronic oscillators[34–36]. Another advantage is the possibility of the combination of the all-magnetic linear read-out with in-memory computing for example magnetic random-access memory[37]. Furthermore, the scalability of this concept to nanoscale dimensions reduces the displacement distances, which would give rise to latencies in the nanosecond regime[17].

This conceptually new reservoir computing idea based on Brownian dynamics of magnetic skyrmions in confined geometries bypasses challenges for existing theoretical proposals for skyrmion RC. Moreover, leveraging the stochastic motion of the skyrmion, allows operation at current densities several orders of magnitude smaller than existing spintronic reservoir computation concepts. The present demonstration is based on a single confined skyrmion, which is already sufficient to realize non-separable operations and the universal set of Boolean functions. Generalizing this concept to multiple confined skyrmions and scalability to nanoscale dimensions suggest a highly promising path to ultralow energy neuromorphic computing.

**Methods**

**Sample parameters**

The thin film layer stack used was sputtered by a Singulus Rotaris magnetron sputtering system, consisting of Ta(5)/Co$_{20}$Fe$_{60}$B$_{20}$(0.95)/Ta(0.09)/MgO(2)/Ta(5) with the thickness of the layer in nanometers in parenthesis[17]. Subscripted numbers are the relative atomic concentration of the respective element in percentage. The sample is specifically tailored for low pinning and skyrmions above room temperature, which exhibit thermal diffusion. Even though this Cobalt-Iron-Boron based thin film inherits a very flat energy landscape[17] and thus low pinning which also enables thermal diffusion of skyrmions, pinning can influence the position of the skyrmion in relation to the corners.

The structures were patterned by electron beam lithography (EBL) using an EBeam Pioneer machine and then etched by Argon ion etching using an IonSys Model 500 ion beam etching system. 15 µm long wires in 120° relation to each other at the corners were used in the layout of the triangle to ensure better electrical connection with the gold pads. The width of the wires is between 2 to 5 µm, the skyrmions used in this work do not enter the wire and stay in the triangle due to the skyrmion edge repulsion. Different device sizes ranging from 22 µm to 40 µm edge length were tested, but for this work a device size of 40 µm was used. For the electrical contacts, a lift-off technique was used after EBL was done for the layout of the pads. The contact pads consist of 5 nm of chromium and 60 nm of gold on top and have a base size of 250×250 µm². Contacting is established using an aluminum wire which is bonded from the pads to the sample's home made holder.

**Measurement setup**

The sample itself is placed on a QC-17-1.0-2.5MS Peltier element to achieve the necessary temperature of around 330 K at ambient air, measured by a Pt100 resistive heat sensor, to realize the skyrmion phase and the size of skyrmions appropriate for the operation. Achievable temperature range was 285-360 K with a temperature stability of 0.3 K. The size of the skyrmions and their thermal diffusion directly depend on the temperature[17]. Albeit necessary for the skyrmions in our stack, the increased temperature also leads to increased diffusion which thus can lead to a necessity of higher current densities to keep a skyrmion in one corner. If the thermal energy is too high, the skyrmion can jump/move towards the middle or other pinning sites in the device, which can be outside of the measured circles or, in the extreme case, annihilate.

We observe the magnetic structures with a commercial Evico GmbH magneto-optical Kerr effect (MOKE) microscope with a CCD camera connected to a PC. The magnetic out-of-plane field is supplied by a self-built electromagnetic coil, while the in-plane magnetic field pulse is supplied by a rotatable electromagnetic coil from the microscopes manufacturer. Images and videos are taken using the polar magneto-optical Kerr-effect, recorded by a Hamamatsu Digital CCD Camera (C8484-03G02) with a CCD spatial resolution of 1344×1024 pixels. The videos are recorded with a framerate of 16 frames per second at an exposure time of 62.5 ms. To achieve this frame rate, we use a 2×2 binning (4 physical pixels are averaged to one virtual pixel in the image) resulting in a resolution of 672×512 pixels with a field of view of 125×95 µm. Differential images between skyrmion and saturation states were used to enhance the contrast. This leads to black and white subtraction errors at the edges of the device in the images (owing to incomplete overlapping), which causes an operation error when the sample/structure moves under the microscope due to thermal drift or mechanical strain. We used a stable thermal equilibration state and increased mechanical stiffness to reduce the drift. Additionally, a script for repositioning the device in the pictures was used to further reduce the thermal and mechanical drift which occurs during recording, thus ensuring more correct tracking.

**Skyrmion imaging and induced motion**

Skyrmions are nucleated by setting an out-of-plane (OOP) field in the µT regime and applying an in-plane pulse field of 35 mT, by switching off the field after it saturated the sample for a second. The amount of skyrmions nucleated in the structure depends on the set OOP field since the radius of the skyrmions can be influenced by the OOP field and the number of skyrmions the geometrical structure can harbor. By increasing the OOP field, the size of the skyrmions is reduced till skyrmions start annihilating and until only one single skyrmion is left. Decreasing the OOP field again to a fixed value led to a reproducible skyrmion size. The skyrmion moves around the center of the device due to thermal motion but is mostly restricted of staying in the center due to the edge repulsion of the device.

The electric potentials are applied by two independent voltage sources (Keithley 2400 SourceMeter), which are connected to the ground through a custom-made breakout box. The latter prevents the device from breaking due to unintentional current flow or induction of current in the wires/cables due to the magnetic fields. The breakout box allows to safely apply the voltage potential to the sample.

Depending on the structure, the potentials for the current-induced skyrmion motion range from 2 mV up to 5.5 mV, which corresponds to a few µA of electrical current. Considering the geometry, the current densities used at the half-width of the triangle ranges from $2\times10^7$ A m$^{-2}$ to $3\times10^8$ A m$^{-2}$. Although the geometry is symmetrical, the resistance between pairs of the corners of the triangle is not the same and varies due to possible sputtering inhomogeneities, the connections of the gold pads, and varying bonding connection quality of the attached wires.

This is compensated by applying different weights in the linear regression, which also takes possible pinning sites in the triangle into account. With increasing current density, the temperature increases in particular at the tips of the triangle element. The increased temperature enhances the motion of the skyrmion and its likelihood of annihilating, if the current density, and thus the temperature, becomes too high.

**Image analysis**

For the analysis of the videos, the skyrmion in the device was tracked using the trackpy package for Python[38,39] and its center position for every frame was compared to the overlaying mask of the four center-symmetric circles on the device. The number of frames, in which the skyrmion resides in a certain circular region are summed up and put into relation to the overall number of frames in the video, resulting in the probability of the skyrmion being in a certain area while a fixed current is applied. Fig. 1 in the supplementary is the heatmap of the probability of a position for a single skyrmion in the device. One can see that the likelihood at certain positions is increased, indicating the existence of pinning sites[40], on which it is energetically favorable for the skyrmion to stay. Thermal activation leads to motion around said pinning sites. This thermal activation can be increased using a higher temperature, until the skyrmion becomes either too small for our experimental setup's resolution or annihilates.

As an outlook, using the three different input types with three physical inputs, 27 possible combinations would be possible, although some combinations have either no function (e.g. [- - -]) or are redundant due to gauge invariance of the electric potential (e.g., state combination a) - - 0 and b) 0 0 +, resulting in the skyrmion to move towards the corner with a) 0 or b) +). When more skyrmions are present in the confinement, the systems' response to the inputs becomes more complex. This is expected to be most prominent for skyrmion numbers incommensurate with the geometry Additionally, combining multiple devices would lead to even more states, thus leading to even higher capacity[37].

**Acknowledgements**

The project was funded by the Deutsche Forschungsgemeinschaft (DFG, German Research Foundation) projects 403502522 (SPP 2137 Skyrmionics), 49741853, and 268565370 (SFB TRR173 projects A01 and B02) as well as TopDyn. The work is a highly interactive collaboration supported by the Horizon 2020 Framework Program of the European Commission under FET-Open grant agreement no. 863155 (s-Nebula) and ERC-2019-SyG no. 856538 (3D MAGiC), the Zeiss foundation through the Centre for Emergent Algorithmic Intelligence and is part of the Shell-NWO/FOM-initiative "Computational sciences for energy research" of Shell and Chemical Sciences, Earth and Life Sciences, Physical Sciences, FOM and STW. M. B. thanks the DFG TRR146 for partial financial support. We would also like to acknowledge helpful discussions with Peter Virnau. T.D. gratefully acknowledges financial support by the Canon Foundation in Europe.

**Competing Interests**

The authors declare no conflict of interest.

**Author Contributions**

M. K. and J. M. devised and supervised the study with P. V.. K. R. designed and fabricated the sample. K. R. and G. B. carried out the measurements with support of T. D.. G. B. evaluated the video data. M. B. and J. R. analyzed the evaluated video data and performed training of the linear read-out. The paper was written by K. R., M. B., M. K. and J. M. All authors commented on the results and contributed to writing the paper.

**Code availability**

The computer codes used for data analysis are available upon reasonable request from the corresponding author.

**Data availability**

The data supporting the findings of this work are available from the corresponding authors upon reasonable request.

## Supplementary Information

**Energy comparison**

To estimate the energy consumption of the device, we assume that we can linearly scale it down into the nanometer regime with an edge length of 50 nm and a skyrmion size of 5 nm (similar ratio of device size and skyrmion size). The read-out would be done using nanometer sized MTJs. We calculate the resistivity of our manufactured device by approximating the geometry between two connections as a wire with the width of half the height of the triangle (17.3 µm) and the length of 3/4 of the triangle's edge length (30.0 µm). The obtained resistivity is then applied for the hypothetical downsized device which results in a resistance of $R \approx 1$ kΩ. Using the lower limit of the current density of $J = 5 \times 10^7$ A m$^{-2}$ would lead to a current flow of 12.5 nA to move a skyrmion. To estimate the time for one operation of the system we use the current-induced motion and diffusion constant values of Ref. [17] at ambient temperature, which uses a similar materials stack. The skyrmion must move around 28 nm (outer radius of the triangle) from the center to one corner. We are considering the current-induced motion velocity of ref. 17 of $v = 5$ mm s$^{-1}$. The resulting time the skyrmion moves towards the corner is $t_c = 5.6$ µs. If the potential is turned off, we use the diffusion coefficient $D = 0.31 \times 10^{-12}\ m^2 s^{-1}$ at ambient temperature from Ref. 17 to calculate the time for the re-set of the device

$$t_d = r^2/4D, \quad \text{(S1)}$$

where $t_d$ and $r$ represent the required time for the re-set operation and the distance between the center and the corner of the device. The calculated time is $t_d = 0.632$ ms. Hence, the operation speed is mostly limited by the re-set process. In this calculated time, we dismissed that the edge repulsion in the tip of the triangle would lead to faster motion towards the center of the device and thus the re-set time is highly over-estimated. The read-out time of MTJs' are in the nanosecond regime and thus negligible compared to the time the skyrmion needs to move. The electrical power needed for this device (MTJs excluded) using Power $P = RI^2$ with the current $I$ = 12.5 nA and resistance of 1 kOhm is $P$ = 170 fW. Considering the calculated times before and one operation of pushing a skyrmion into a corner and resetting the system into the ground state, the energy per oscillation/operation is $E = Pt_{c+d} = 107.4\ aJ$. Compared to the downscaled version of the spin-torque oscillator of Ref. [34] of hundreds of Attojoule per oscillation, the calculated energy value seems viable, despite the rough assumptions.

**Necessity of thermal diffusion**

The functionality of our device is dependent on the stochastic thermal motion, as it is necessary to make certain combinations of input distinguishable, for example ground on top and positive potentials on the bottom two contacts [0 + +]. Since the skyrmion can be at either corner of the applied positive potential, the thermal motion causes the skyrmion to be able to switch between both corners. Without stochastic dynamics, the skyrmion could be pushed into one corner and be unable to leave this position, the systems' response to, for example, the [0 + +] input combination would not differ from of the response of either [0 0 +] or [0 + 0].

**Discussion and interpretation of the weights for linear read-out**

As the weights are optimized, the vector of local skyrmion occurrence probabilities in the four regions, which is a consequence of the applied input voltages, is mapped to either 0 or 1 depending on the desired output for the given Boolean operation for each input. Therefore, the values weights in Table 1 can be understood taking into account the global skyrmion occurrence distribution in Sub. Fig. 1. Consider the OR operation: The output should be 1 for input-voltages [0 mV, 2 mV], [2 mV, 0 mV] and [2 mV, 2 mV] at the [left, right] corners of the triangle and 0 for [0 mV, 0 mV]. In the former cases, occurrence is always high in the left and/or right region and therefore the weight is positive, whereas occurrence in the top region indicates a [0 mV, 0 mV] input which should be mapped to 0 and therefore the weight for the top region is negative. In principle, since the results of the Boolean operations are invariant under an input transformation [A, B] to [B, A] and since the read-out regions are placed symmetrically around the vertical mirror axis of the triangle, the weights for the left and right regions should be identical. However, due to asymmetries in the resistance between the contacts and pinning effects, they are not. This does not hinder the functionality of our device since the weights automatically adjust for these effects. Moreover, note that the weights (except for the intercept) for NOT operations are just the negatives of their counterpart, as the not operation only exchanges if the result should be mapped to 0 or 1.

Since the [0 mV, 0 mV] input combination is the only one generating high local occurrence probability in the top region (Sup. Fig. 1), the operations which only need to separate this input from the other ones (i.e., OR and NOR) show reduced noise as seen in Fig. 3.

**Current density distribution**

To understand why and where skyrmions are annihilated as the current is increased, a minimalistic COMSOL Multiphysics®[34] simulation has been performed. Only the 5 nm top tantalum layer is simulated, and the gold/chromium contact is modelled as a chromium pad enclosing the sides (but not the top) of the rectangular extensions at the corners. Experimentally, we observed that with a higher current density, the skyrmion would first get pushed into the corner and then get pulled into the high current spots in the simulation in supplementary Fig. 2. The probability of a skyrmion annihilating was highest.

**Additional References**
34. COMSOL Multiphysics® v. 5.5. www.comsol.com. COMSOL AB, Stockholm, Sweden.

**Supplementary figures and graphs**

- Supplementary Table 1: weights of the device
- Supplementary Figure 1: heat map
- Supplementary Figure 2: current distribution simulation
- Supplementary Figure 3: multiple skyrmion states

**Supplementary Table 1| Weights of linear regression**

| Operator | $W_{\text{left}}$ | $W_{\text{right}}$ | $W_{\text{top}}$ | $W_{\text{middle}}$ | $W_{\text{intercept}}$ |
|---|---|---|---|---|---|
| AND | 1.124 | 7.497 | 0.623 | 1.022 | -0.797 |
| NAND | -1.124 | -7.497 | -0.623 | -1.022 | 1.797 |
| OR | 0.272 | 1.09 | -1.403 | -2.908 | 0.942 |
| NOR | -0.272 | -1.09 | 1.403 | 2.908 | 0.058 |
| XOR | -0.852 | -6.407 | -2.027 | -3.93 | 1.739 |
| XNOR | 0.852 | 6.407 | 2.027 | 3.93 | -0.739 |

**Supplementary Table 1|** Weights of the read-out for the different positions of the skyrmion for the exemplary Boolean operations.

**Supplementary Figure 1| Heat maps**

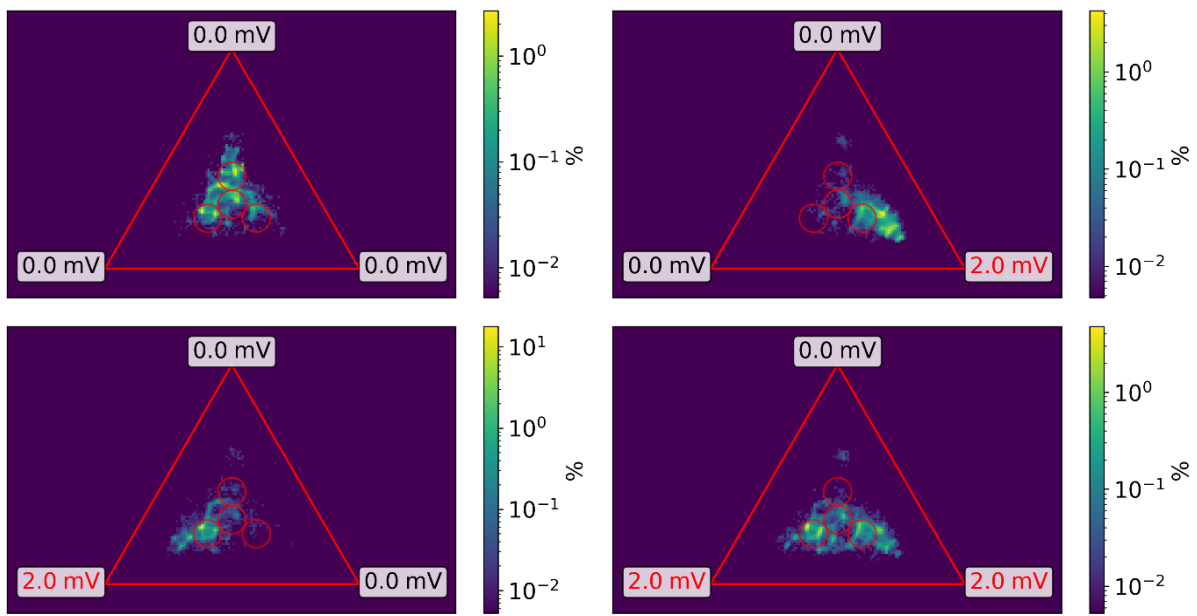

**Supplementary Figure 1|** Heat maps of the device showing the skyrmion occurrence probability. Yellow represents highest likelihood of a skyrmion position. In the corners the respective, applied potential is shown in either black for the ground potential (0.0 mV) or in red for +2.0 mV potential. a) Free diffusion, b) skyrmion is pushed to the right, c) skyrmion is

pushed to the left and d) skyrmion is pushed to the left and right. Red circles are the read-out area used for the linear read-out.

**Supplementary Figure 2| Current density distribution simulation**

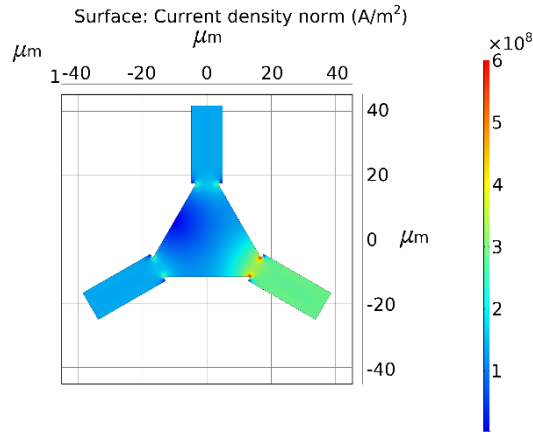

**Supplementary Figure 2|** Minimalistic COMSOL Multiphysics® simulation of the current density distribution in our device. Left: the current density distribution for one positive potential applied (right bottom) and two grounds. Right: current density distribution for one positive potential (right bottom) one ground (top) and a floating connection (left bottom). The points of highest current density are the (red) bends where the gold contact ends on the stack. Increased current leads to higher temperature here which can result in the annihilation of skyrmion at these points.

**Supplementary Figure 3| Multiple skyrmions**

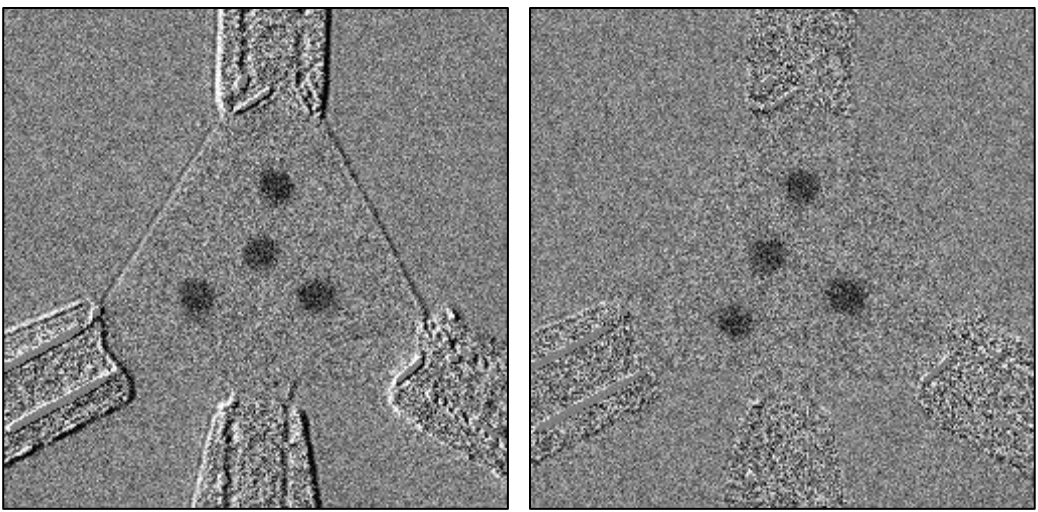

**Supplementary Figure 3|** Multiple skyrmion states in the device. By using multiple skyrmions the amount of states and thus the capacity of the device increases.